# On the decoupling of relaxation modes in a molecular liquid caused by isothermal introduction of 2nm structural inhomogeneities.

Kazuhide Ueno and C. Austen Angell,
Dept. of Chemistry and Biochemistry,
Arizona State University, Tempe, AZ 85287-1604

**Abstract**

**To support a new interpretation of the origin of the dynamic heterogeneity observed pervasively in fragile liquids as they approach their glass transition temperatures $T_g$, we demonstrate that the introduction of ~2 nm structural inhomogeneities into a homogeneous glassformer leads to a decoupling of diffusion from viscosity similar to that observed during the cooling of orthoterphenyl OTP below $T_A$, where Arrhenius behavior is lost. Further, the decoupling effect grows stronger as temperature decreases (and viscosity increases). The liquid is cresol and the ~2nm inhomogeneities are cresol-soluble asymmetric derivatized tetrasiloxy-based (POSS) molecules. The decoupling is the phenomenon predicted by Onsager in discussing the approach to a liquid-liquid phase separation with decreasing temperature. In the present case the observations support the notion of a polyamorphic transition in fragile liquids that is hidden below the glass transition. A similar decoupling can be expected as a globular protein is dissolved in dilute aqueous solutions or in protic ionic liquids.**

**Introduction**

One of the most intensively researched aspects of glassforming liquids is the heterogeneity that develops in their dynamic properties as temperature is decreased from above the melting point to deep in the supercooled liquid state[1-6 7,8]. It is found that, except for the case of polymer liquids, the behavior at high temperatures is simple: fluctuations, such as those explored by neutron scattering through the self-intermediate scattering function, relax exponentially on time scales that follow the simple Arrhenius equation. Then at temperatures that are usually, but not necessarily, below the melting point, (i) a shoulder develops in the relaxation function, (ii) the longer time component of the relaxation function becomes non-exponential and (iii) the temperature dependence of the relaxation time (both average and most probable) departs from Arrhenius behavior[9]. The departure from Arrhenius behavior can be spectacular in the case of many liquids, now designated as "fragile". It is usually also these liquids whose relaxation functions depart most rapidly from exponentiality on temperature decrease, though this latter relationship is not always found.

The origin of these features is one of the most vexing of the unanswered questions in this research area, and has become the source of many alternative theoretical ideas, the most influential of which has undoubtedly been the mode coupling theory MCT[10-12]. MCT has been very successful in predicting the details of the two-step transition from simple exponential behavior at high temperatures to non-exponential behavior at lower temperatures though it is unable, by its mean field nature, to describe the lower temperature, activated, features of the relaxation time behavior (near $T_g$).

An additional, and closely related, characteristic of this low temperature region of liquid behavior that has been given considerable attention is the decoupling of diffusion from viscosity that is observed to accompany the departure from exponential relaxation[9,13-17]. The decoupling is also observed for translation relative to rotation, the translational motion being less affected by decrease of temperature than either the rotation or the shear relaxation[7,89]. The decoupling is often assumed to set in at the mode coupling theory critical temperature, $T_c$(MCT) which frequently coincides with the $\alpha-\beta$ bifurcation temperature observed in dielectric relaxation. However, when precise data are examined it is seen to originate at temperatures closer to that at which the shoulder in the relaxation function develops. In molecular dynamics studies, which until recently have been conducted almost always on time scales shorter than that at $T_c$, the development of heterogeneous dynamics is much in evidence above $T_c$ and the correlated "string-like" motions of the fastest 5% of the particles have been studied in much detail[6,18,19].

An outstanding example of the decoupling phenomenon is that of orthoterphenyl (OTP) in which the diffusivity has been measured down to $T_g$ by Ediger and coworkers[16]. Whereas in "strong" liquids, the diffusivity of the slowest particles reaches $10^{-22} m^2 s^{-1}$ at $T_g$[20], in OTP the value observed at $T_g$ is only $10^{-20} m^2 s^{-1}$. This corresponds to a decoupling by two orders of magnitude. It is also two orders of magnitude greater than the value expected from the viscosity at this temperature by the usual Stokes-Einstein scaling. It even continues to increase after the dispersion in relaxation times (non-exponentiality) has stopped changing, apparently eliminating the spectral broadening as an explanation, at least for OTP[8].

It has been common to view the heterogeneity as a feature of the dynamics alone, based on the general failure to observe any corresponding structural features[7], as well as on theoretical expectations[21,22]. The heterogeneity has been characterized by NMR measurements in terms of a dynamic length scale which seems to be of the order 2 nm[23]. This length is considered to describe the size of domains in which the particles move more slowly than in intervening "fast" domains. This same length scale is also implied by the photobleaching experiments of probe molecules of different dimensions[3]. Probe molecules of dimensions larger than about 2 nm are observed to relax exponentially (implying that any heterogeneities in the mobility on this length scale are being averaged out[3,24]. Accordingly, probe molecules of dimensions approaching that of the host molecule relax with degrees of non-exponentiality that approach that characteristic of the host molecules[25]. This is consistent with the observation of Kawasaki and Tanaka[26], by MD studies, that a system appears non-ergodic unless sufficient time is allowed for the heterogeneity length scale to be equilibrated, a time scale much longer than that assigned to the $\alpha$ relaxation.

On the other hand, the 2 nm dimension has been attributed to fluctuations in the dielectric constant of glassformers near their $T_g$, based on an excess light scattering observed on cycling through the glass transformation zone[27]. This would imply that density variations of this magnitude are being detected. This in turn would imply that the heterogeneity observed might have a structural origin that is simply more sensitively detected in dynamic measurements (relaxation spectroscopy being the most structure-sensitive of all spectroscopies[28]). Indeed this idea has been supported by the recent simulations of Harrowell and coworkers[29,30] who use a clever method of separating the dynamics from the structure in such a way as to enhance any correlation between the structural organization and the subsequent particle dynamics,. They detected a "propensity" for fast dynamics that traces back to structural features.

In 1945, Onsager[31] anticipated that composition fluctuations in systems with large positive deviations from Raoult's law (and critical solution points at lower temperature) would lead to increases in viscosity because of the interruption of the flow lines controlling viscosity, while diffusivity would be little affected. Onsager's line of thought (which was quickly verified) should be equally valid for structural fluctuations anticipating an isocompositional (polyamorphic) phase transition such as that suggested for fragile liquids by Matyushov and one of the present authors[32], and for unusual systems like water and silicon[33] in the liquid state. This is the essence of the idea that we wish to explore in this paper.

We take a simple approach to exploration of this problem. Rather than waiting for the liquid to generate structural heterogeneities of the 2nm dimension by sufficient cooling, we introduce the heterogeneities by direct addition of a soluble component with core size of the 2 nm dimension. For this study, a small-molecule solvent is chosen such that the study is conducted above $T_c$ (for the solvent). This, and the fact that the solvent, m-cresol, is also non-fragile in its behavior in the temperature of the study, guarantee that (like the non-fragile liquid glycerol [sec.1.9.1 of ref.[9]] and non-fragile polymer polyisobutylene[34-36]) it is free of any heterogeneities of its own generation. Then we study the relative changes of viscosity and diffusion that occur as result of the addition. If the flow lines are indeed interrupted by the nanoscopic inhomogeneities, then the viscosity will increase much more rapidly than will the diffusivity.

Diffusivity is difficult to measure with high precision, particularly in condensed phases, and the difficulty increases with decreasing diffusivity. The diffusion/viscosity decoupling on approach to critical points and spinodal lines has been more precisely demonstrated using the (precisely measurable) equivalent conductivity in place of diffusivity[37] (to which it is related by the Nernst-Einstein equation). We therefore adopt that strategy in the present study.

For a solute of the appropriate size we have chosen one of the tetrasiloxy family generally known as POSS molecules[38], the particular choice being that of aminopropylisooctyl POSS in which there is pendant amine in one of the possible alkyl group sites (see insert to Fig. 1). This molecule yields a viscous liquid at ambient temperature. The density of the liquid at 20ºC is 0.99, corresponding to a molar volume of 1280 ml/mole from which we deduce a volume per molecule of 2.13 nm$^3$, and accordingly a diameter of about 1.6 nm depending on how we assume the molecules to pack. We will refer to the POSS molecules as "~2nm inhomogeneities". Notwithstanding the high viscosity, the glass temperature of POSS is quite low, -53.3ºC (according to the DSC studies at 20K/min reported below). While this molecule has no conductivity of it own, it can be made conducting by protonation of the amine group, using a variety of acids. Alternatively, and preferably, ions of the same size as the molecular solvent can be introduced as a separate dilute component.

We have studied two cases of the latter method of introducing conductivity, choosing in each case m-cresol ($T_g$ = -77ºC) as the non-fragile molecular liquid. In the first, we use m-toluidinium methanesulfonate as the matching salt. This salt is the protic ionic liquid made by adding methanesulfonic acid ($CH_3SO_3H$) to toluidine. M-Toluidene, like m-cresol, is a disubstituted benzene, with methyl and amine groups attached to the ring in meta conformation. Thus the only difference between the cation of our ionic dopant, and the host solvent is the replacement of the –OH group of cresol by the more basic –$NH_2$ group. This group becomes protonated to give –$NH_3^{+.}$ The methanesulfonate anion $CH_3SO_3^-$ is comparable in size to the cation, and the solvent molecule, hence their motions would seem to provide a reasonable guide to the motions of the solvent. In the second case we use an *aprotic* salt to avoid a possible source of uncertainty in our results described in the following.

An uncertainty could arise because the POSS molecule itself contains a pendant amine group which could be protonated at the expense of the toluidine, and thereby affect the ionic distribution, in particular the ion-pair concentration. To eliminate this concern we have also made measurements using the aprotic salt, butylmethylimidazolium tetrafluoroborate (BMIm $BF_4$) [$BMIm^+$]$BF_4^-$] as the source of conductivity. The distilled under vacuum prior to use. 1-methylimidazole from Alfa Aesar, $NaBF_4$ and methanesulfonic acid from Aldrich were used as received. The aprotic salt, 1-butyl-3-methyl imidazolium tetrafluoroborate (BMIm $BF_4$), was prepared following a slight modification of the procedure reported by Welton and co-workers[39] and dried under vacuum at 70°C. The protic salt, m-toluidinium methanesulfonate, was formed by neutralization of the pure Brönsted acid with pure base. An equimolar amount of m-toluidine and methanesulfonic acid were reacted together using an ice bath to control temperature. The product was washed with acetone several times and dried under vacuum.

10mM electrolyte solutions in m-cresol were first prepared and these solutions then mixed with POSS to the stated final compositions which are reported in wt % POSS. A solution that is 10 wt % POSS is 10.5 volume% POSS and 0.45 mol % POSS. Only the volume % is physically relevant to our problem. Since the POSS molecule is roughly spherical in shape, it is estimated that the inhomogeneities will percolate at about 30 volume % (~29 wt % POSS). We are interested in the domain well below percolation.

Viscosity measurements were performed using a Brookfield cone-plate viscometer (RVTDCP, CP-40) the temperature of which was controlled to 0.1ºC by means of a water circulating bath (LC20, Lauda). The instrument was calibrated at Brookfield Co. and the accuracy was confirmed to 0.5% by measurement on a viscosity equivalent conductivity for this solution proves to be higher than with the toluidinium methanesulfonate, probably as a result of being more completely dissociated. Both solutions give qualitatively similar behavior on addition of POSS, as will be seen below, so both support our expectations.

**Experimental section.**

The POSS molecule for our study, aminopropylisooctyl POSS (AM270, MW = 1267) was obtained from Hybrid Plastics and was used as received. The solvent for the study, m-cresol, and the ion precursor, m-toluidine, were purchased from Aldrich and were standard (Dow 200 fluid, 100cSt). Ionic conductivities, in the range of $10^{-5}$-$10^{-6}$/cm, were determined from complex impedance data obtained using a Solartron 1250 frequency response analyzer with a frequency range of 10Hz to 65kHz. The solution was contained in a dip-type cell with platinum electrodes, the temperature of which was controlled to 0.1ºC by a Peltier temperature controller. The cell constant, of value 0.59 cm$^{-1}$, was determined using a 0.01M KCl aqueous solution.

**Results**

Viscosities for cresol and for its solutions with POSS between 0 and 30 wt%, are shown in Figure 1 as a function of temperature, using the Arrhenius presentation of data. The "**m** fragility"[40] of m-cresol is reported as ~65 (D = 10)[41], based on viscosity data. The **m** fragility of pure POSS AM270, according to the variable scan rate differential scanning calorimetry protocol developed by Velikov[42] and Wang[43], is 76, which is comparable to the value of 74 reported by Kopesky et al for a methacrylate POSS[44]. The conductivities likewise show non-Arrhenius temperature dependences, though we do not consider these dependences here, beyond noting that, in the temperature range of interest, they are the same as for viscosity when no POSS is present. VFT parameters for the small temperature range (only 1.5 orders of magnitude) are not suitable for determining **m**-fragility, though the VFT equation fits the data well. As in many cases of high temperature data on aromatic molecular liquids[45,46], they VFT equation fits imply (vanishing mobility temperatures) $T_0$ in excess of $T_g$. Thus **m** values (which refer to behavior near $T_g$) are not assignable. In any case, they are not of importance to this study.

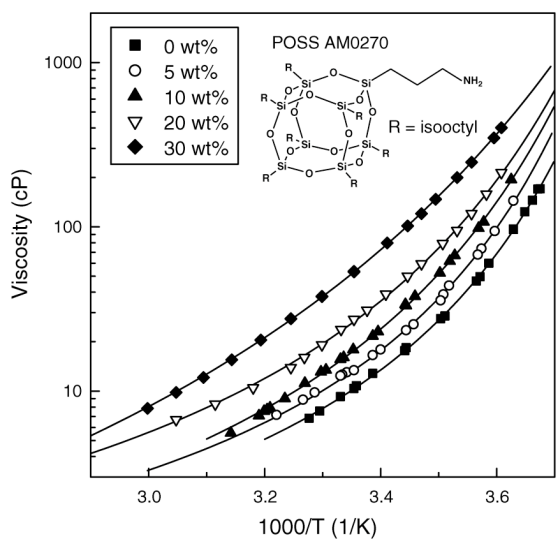

**Figure 1.** Viscosity of m-cresol and its solutions up to 30 wt % POSS, presented as Arrhenius functions of temperature.

We therefore address our original motivation, which was to compare the rate at which the viscosity changes, as the bulky POSS entities are incorporated into the solvent medium, relative to the rate at which the solvent mobility is affected. As marker for the latter we are using the ionic mobilities of the ionic dopant. Accordingly, we present the conductivity data as the value of the conductivity relative to the conductivity prior to the addition of any POSS. This is presented for both specific and equivalent conductivity ratios, i.e. $\sigma_0/\sigma$ and $\Lambda_0/\Lambda$, in Figure 2. The relevant concentration range is that in which the POSS is the minority component so that we can validly consider that we are observing, primarily, the effect of introducing ~2nm structural inhomogeneities into the milieu of the mobile solvent molecules. At large enough POSS concentrations (> ~30 vol%, or 31 wt %) the inhomogeneities will percolate and a different behavior, not of interest to the present purpose, will be encountered. To avoid obscuring the important data, we present the findings for conductivity due to the protic salt and that due to the aprotic salt, in separate figures, taking the latter unambiguous case first. It is clearly seen that the addition of inhomogeneities at 25ºC barely affects the conductivity up to 10 wt% (9.7 vol %) POSS while it has a marked effect on the viscosity.

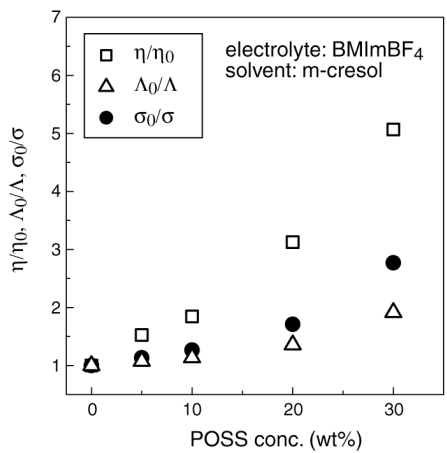

**Figure 2.** Comparison of the relative rates of change of viscosity and conductivity as POSS is added to the initial cresol solution of BMIBF$_4$. Note how the viscosity begins to rise immediately while the conductivity, either specific or equivalent, is initially little affected by the additions. The temperature of the study is 25ºC, at which the viscosity of the initial cresol solution is 10.3 mPa.s.

The data for the effect of POSS additions on the solution with *protic* salt additions, are shown in Figure 3. On a relative conductivity basis there is little difference to be seen between the two cases. Quantitatively, the conductance obtained with the aprotic salt is a little higher than for the protic salt, presumably due to a smaller extent of ion pairing. This effect may be responsible for the small initial increase in conductivity in this case, an effect not seen with the aprotic salt solution.

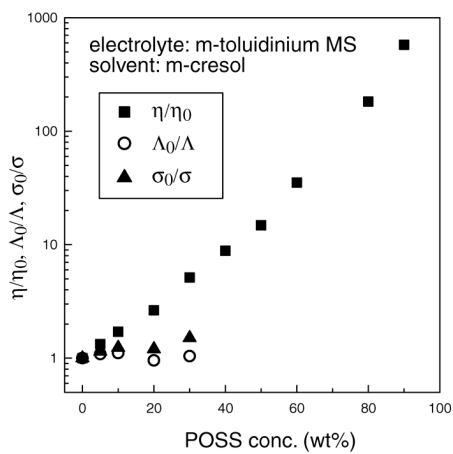

**Figure 3**. Comparison of the relative rates of change of viscosity and conductivity as POSS is added to initial cresol solution of the protic salt, toluidinium methanesulfonate. Temperature is 25ºC.

**Discussion.**

Figure 2 shows very clearly the effect that Onsager predicted would accompany the introduction of physical inhomogeneities as a system approaches a critical solution point in a binary solution with immiscibility. At 25ºC, where the viscosity is only 10.3 mPa.s, the effect on the viscosity of adding 10 wt% (9.7 vol%) of inhomogeneities amounts to almost a factor of two. The change of conductivity in the same composition range is very small. A doubling of the viscosity without change of solvent mobility may not seem much compared with the two orders of magnitude observed between $T_A$ and $T_g$ in fragile o-terphenyl but it must be remembered we are examining the effect at low viscosity. In the Einstein theory for effect of blockers on viscosity, the energy dissipation increases proportional to the viscosity[19], so the small effect we see for 10 % inhomgeneity volume at solvent viscosity = 10.3 mPa.s would inflate during cooling to $T_g$. would be greatly magnified with decreasing temperature (see below). Thus one can readily see how the generation of ~2nm structural inhomogeneities during cooling would cause the observed decoupling. The intrinsic viscous slowdown due to change of temperature on a homogeneous medium would be augmented by the inhomogeneity effect, making the liquid look more fragile in the viscosity measurement.

As originally noted by McCall et al.[47] for the case of o-terphenyl, the diffusivity and viscosity adhere to the Stokes-Einstein relation over very wide ranges of viscosity within a small factor - indeed it is only in the last 20 K or so (~$0.1T_g$) that the deviation develops strongly. This is emphasized in the plot of the log (deviation) vs. $T/T_g$ given by Mapes et al.[16] in their figure 3, which shows data for OTP by different techniques fitted by two different theoretical approaches[48,49], and Figure 4 which shows data for different polyphenyl liquids. Neither of the theoretical approaches considers a first order phase change to be imminent, thus the interpretation being supported here is of novel character. In the analysis of Matyushov[32], the first order transition terminating the high temperature liquid state (under ergodic conditions) has been predicted to lie some 10% below $T_g$, and a spinodal (at which the enthalpy fluctuations would diverge) lies just a few degrees lower.

We noted above that according to Einstein the decoupling we observe, due to viscosity increase in excess of the decrease in mobility, should increase with increasing viscosity. We can observe this effect for a small range of temperatures with the present data simply by repeating the analysis of Figures 2 and 3 at the higher temperature of 35ºC. This is shown in Figure 4, for the aprotic salt case. It can be seen that the effect of POSS additions on the viscosity diminishes significantly as the temperature increases (and background viscosity decreases). The effect of the temperature change on relative conductivity, on the other hand, is minor. Here the viscosity change is only 40 %, which is very small compared with the 15 orders of magnitude change typical between melting point and $T_g$. Thus it can be appreciated that large decouplings can be provoked by the development of a relatively small volume fraction of structural inhomogeneities as the system approaches its glass transition temperature.

**Diffusivity-viscosity decoupling in pure liquids near a spinodal**

For a case in which the liquid-liquid critical temperature, for reasons peculiar to the system, rises above the glass temperature, pre-critical inhomogeneities should be generated at temperatures that correspond to much lower viscosities - and then the Stokes Einstein equation will be observed to break down at much lower viscosities. A case in question is that of water, in which the second critical point[48,50,51] (or perhaps the liquid-liquid coexistence line[52] of which the critical point is the limit), lies at 220-230K (while $T_g$ lies much lower at 136-200K depending on the state of confinement of the system.

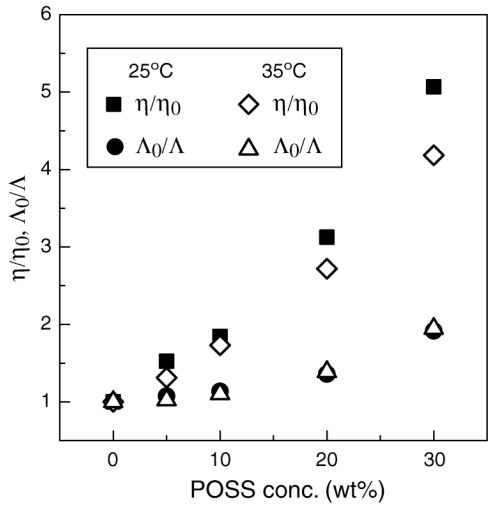

**Figure 4.** Effect of the concentration of ~2 nm inhomogenities (i.e. wt % POSS) on the relation between viscosity and conductance in cresol doped with the aprotic salt [BMIm$^+$][BF$_4^-$], at two different temperatures (hence two different initial viscosities). Temperature increase is seen to diminish the effect of 2nm inhomogeneities on the viscosity while the effect on mobility (equivalent conductivity) is unaffected. Thus the decoupling effect of nanoinhomogenieites grows stronger as temperature decreases.

The S-E breakdown in water has been under discussion since the diffusivity was measured to -33ºC in 1972[53]. A recent summary and analysis of best data[54] suggests that the crossover to the fractional Stokes-Einstein equation occurs around 260K - where the viscosity is only ~$10^{-2}$ Pa.s. Data on computer-simulated water have been used to discuss the independence of breakdown signatures on structural features and populations[55] while elsewhere[56], using two related models with liquid-liquid critical points, the breakdown has been associated with the development of detectable populations of structurally distinct units (see also[57]). In each case, however, a close connection to impending criticality was supported.

Evidently the breakdown of the Stokes-Einstein relation can be induced in different ways. With impending criticality, fluctuations occur on all length-scales up to a maximum value, the correlation length, which itself increases with decreasing distance in temperature or pressure from $T_c, P_c$. In our case we have introduced a single length scale, of value ~2 nm based on the evidence obtained near $T_g$ from both static and dynamic experiments[7,27]. The S-E deviation for such a case conforms more closely to that described in the "obstruction" theory of Douglas and Leporini[58] – which in turn relates closely to the original Einstein treatment of viscosity. While we could, in principle, investigate the effect of introducing inhomogeneities of different nanoscopic length scales, we feel that our purpose, viz., demonstrating that the dispersion effects and their consequences on the diffusivity-viscosity relations seen in viscous liquids on approach to $T_g$ may have a static structural source, has been sufficiently well served by the present data. It would be consistent with what we know from the case of water, seen in other studies as a Rosetta stone for understanding subtleties of glassformers[59], that this origin lies in a hidden criticality. Proving that the structural heterogeneity owes its origin to a liquid-liquid transition hidden below $T_g$ is, however, probably better tackled via an expansion of the current vapor deposition routes[60,61] to novel glassy states not accessible by the normal steady cooling methods[62,63]. We note the evidence[61,64,65] that these novel low enthalpy states are obtained in greatest yield when the deposit is made at a temperature about 10% below the normal $T_g$, and particularly the evidence[64,65] that these low enthalpy materials seem to return to the normal viscous liquid state by a nucleation and growth mechanism, the hallmark of a first order transition.

A natural concern with our proposal would arise from the general feeling that the effects of criticality are only seen very close to the critical point. But we must emphasize that we are by no means asserting that viscous liquids showing diffusivity/viscosity decoupling are approaching a critical point head-on. This would only occur in a very unusual case. It is important to recognize that the effect of fluctuations that arise from the existence of a critical point are felt over a much wider range of temperature when the isobar under consideration is "off-critical". This effect is well illustrated in the heat capacity isobars for the attractive Jagla model[63,66] which has an unambiguous liquid-liquid critical point in the stable liquid range of the system. It can be seen in Fig. 6 of ref.[67] and Figure 9a of ref.[68], that the further from the critical pressure the isobar under consideration lies, the wider the temperature range over which the enthalpy fluctuations associated with the critical point contribute to the excess heat capacity. Since the validity of the Adam-Gibbs equation[69], (which has been broadly validated for diffusivity by model liquid studies[70,71]) implies that enthalpy fluctuations are determining liquid state dynamics in viscous liquids, it is reasonable to conclude that diffusion/viscosity decoupling will also be observable over a wide range of temperatures when the isobar under consideration is off-critical.

**Concluding remarks.**

While we have used the tetrasiloxy-based molecules as our source of ~2 nm structural inhomogeneities for this study, there are alternative possibilities that are not without interest. For instance, the common globular protein lysozyme has a diameter of 3.2 nm[72] and dissolves readily in both water and hydrated ionic liquids. In water, globular proteins tend to aggregate and precipitate at quite low concentrations but in neutral protic ionic liquids, the solubilities can be very high[73]. There are non-fragile ionic liquids like the dihydrogen phosphates of ref.[74] in which the lysozyme could be dissolved to permit a study like the present one. This case would have the advantage that the ions are intrinsic to the system under study. We will report findings on such cases separately.

**Acknowledgements.** K. Ueno has benefited from a fellowship of the Japanese Society for Promotion of Science.